# Elucidating the thermal spike effect by using a coupled classical oscillators model

*Jiajian Guan*,[1,*] *Yue He*,[1] *Bin Liao*,[2] and *Xu Zhang* [2]

[1] Department of Chemical and Materials Engineering, Faculty of Engineering, the University of Auckland, PB 92019 Auckland, New Zealand

[2] Key Laboratory of Beam Technology and Materials Modification of Ministry of Education, College of Nuclear Science and Technology, Beijing Normal University, Beijing 100875, China

*Authors to whom correspondence should be addressed: jgua850@aucklanduni.ac.nz

**ABSTRACT**. Atomic heating is a fundamental phenomenon governed by the thermal spike effect during energetic deposition. This work presented another insight into thermal spike using a coupled classical oscillator model instead of a typical heat diffusion model. The temperature profile of deposited atoms was replaced by oscillator amplitude as an energy descriptor. Solving associated partial differential equations (PDEs) suggests the efficiency of energy transfer from the coupled "hot" to "cold" oscillators essentially relies on the atomic distance *r* and the spring constant *k*. The solution towards the damped wave equation further emphasizes that a relatively low wave speed *v* among coupled oscillators will spontaneously contribute to a localized thermal fluctuation during energy propagation.

## I. INTRODUCTION

Rapid heating and quenching at the atomic scale is typically a fundamental yet crucial phenomenon occurring in the scenario of heavy atoms colliding with matters within exceedingly short periods [1], especially in the processes of energetic deposition [2,3], for instance, filtered cathodic vacuum arc (FCVA) deposition and high power impulse magnetron sputtering (HiPIMS), where swift metal ions with energy order of $10^{-2}$ to $10^{-1}$ keV bombard on substrate surfaces instantly, generating extremely high temperature along the ion trajectory within the time order of picoseconds (ps), and consequently contributing to subsequent physical processes like nucleation, phase transition, and so forth [4-7]. Transient energy transfer from swift high-energy ions (e.g., MeV or GeV) to target materials along the ion tracks involves the processes of ion-electron, electron-electron, and electron-phonon couplings [5], among which, the electron-phonon coupling prominently contributes to rapid heating by transferring electron energies to surrounding lattice atoms due to electronic stopping effect, resulting in transient thermal spike and shock wave [5,8]; nevertheless, low-energy ions (e.g. < keV) induced thermal spike mainly comes from a direct atom-atom interaction dominated by the nuclear rather than the electronic stopping effect [9]. The notion of an as-proposed thermal spike was first introduced by F. Desauer [10], and F. Seitz and J. S. Koehler reconsidered it for metals in cylindrical geometry based on electron-phonon coupling, which is known as an inelastic thermal spike (i-TS) model [11]. G. Szenes proposed an analytical thermal spike (a-TS) model by solving the partial differential equations (PDEs) from the i-TS model [12]. P. Sigmund and C. Claussen used an elastic collision model to discuss the energy contribution of low-energy ions governed by nuclear stopping [3]. According to R. C. Vilão et al.'s study [13], one can obtain a time and radial distance dependence expression of temperature profile by solving a typical heat diffusion equation (1)

$$c_v \frac{\partial T}{\partial t} = \kappa \nabla^2 T \tag{1}$$

$$T(r,t) = \frac{Q}{8\pi^{3/2} c\rho (Dt)^{3/2}} \exp(-\frac{r^2}{4Dt}) \tag{2}$$

$c_v$ and $\kappa$ are the volumetric specific heat and the thermal conductivity, respectively. The general solution to equation (1) is given by the formula (2), where $Q$ is the heat input, $D$ is the thermal diffusivity with $D=c_v/\kappa$, and $\rho$ is the density of the material. In the i-TS model, the energy exchange between electrons and atoms is quantitatively described by equations (3) and (4), here $C_e$ and $C_a$ are the specific heat coefficient for the electrons and atoms, $K_e$ and $K_a$ are the thermal conductivities of the electrons and the atoms, $g$ is the electron-phonon coupling parameter, and $A$ represents the energy distribution [1,14,15].

$$C_e \frac{\partial T_e}{\partial r} = \frac{1}{r}\frac{\partial}{\partial r}[rK_e \frac{\partial T_e}{\partial r}] - g(T_e - T_a) + A(r,t) \tag{3}$$

$$C_a \frac{\partial T_a}{\partial r} = \frac{1}{r}\frac{\partial}{\partial r}[rK_a \frac{\partial T_a}{\partial r}] + g(T_e - T_a) \tag{4}$$

No matter the a-TS or the i-TS model, the descriptions of the thermal spike are essentially based on the temperature profiles derived from the ordinary heat diffusion equation. However, it's unlikely to obtain an exact scenario of energy transfer at the atomic scale by solving the heat diffusion equation for the following reasons:

(i) the temperature descriptor is statistically defined to describe the system with a large number of atoms rather than individual atomic behavior [16],

(ii) the heat diffusion equation fails to unveil the nonequilibrium state where the temperature gradient exists over a length scale smaller than the phonon mean free paths [17,18]. In terms of energetic deposition, it is a nonequilibrium thermodynamics process at the atomic scale [19], and nuclear stopping mainly dominates the energy loss of deposited atoms.

In this work, we dedicate ourselves to presenting a profound understanding of the energy exchange and propagation at the atomic scale under low-energy ion bombardment using coupled classical oscillator model, in which oscillator amplitude is proposed instead of temperature profile as the energy descriptor.

## II. RESULTS AND DISCUSSION

To reproduce the process of low-energy ion bombardment at the atomic scale, we first performed a molecular dynamic (MD) simulation by using the LAMMPS package. As presented in Fig. 2, an energetic Cu atom (20 eV) collides on the Cu (111) surface under the NVE ensemble. Fig. 3 records the evolution of atomic velocity and kinetic energy within picoseconds after collision, demonstrating the average heat diffusion model isn't strong enough to describe the atomic thermal distribution since the discontinuous energy spikes occur along the atom trajectory. Here, we use the coupled classical oscillator model to give a more precise explanation.

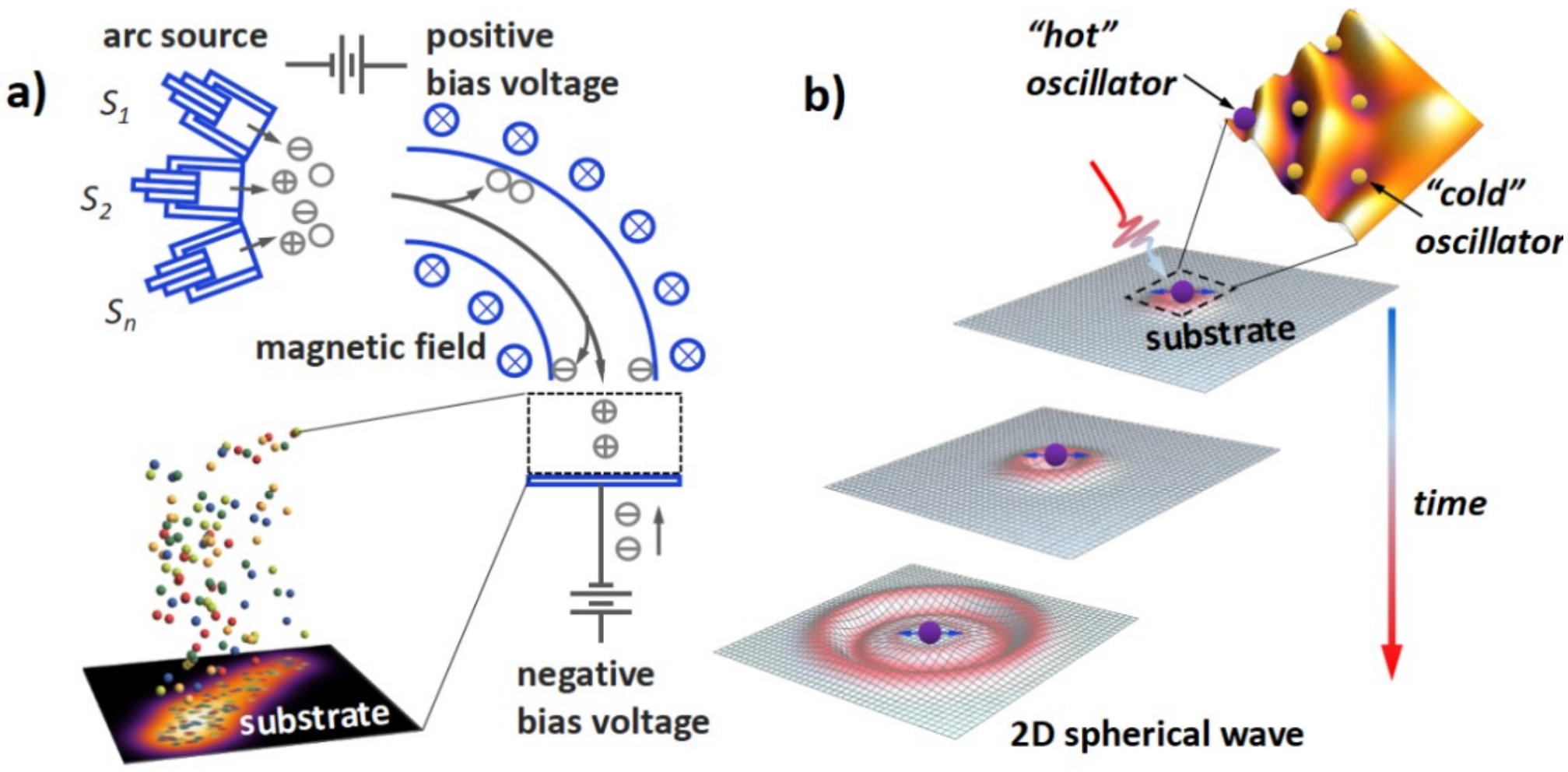

**Figure 1**. (a) Schematic diagram of energetic deposition by using FCVA technique; (b) illustration of time and distance dependence energy propagation from the "hot" to "cold" oscillators as a classical spherical wave.

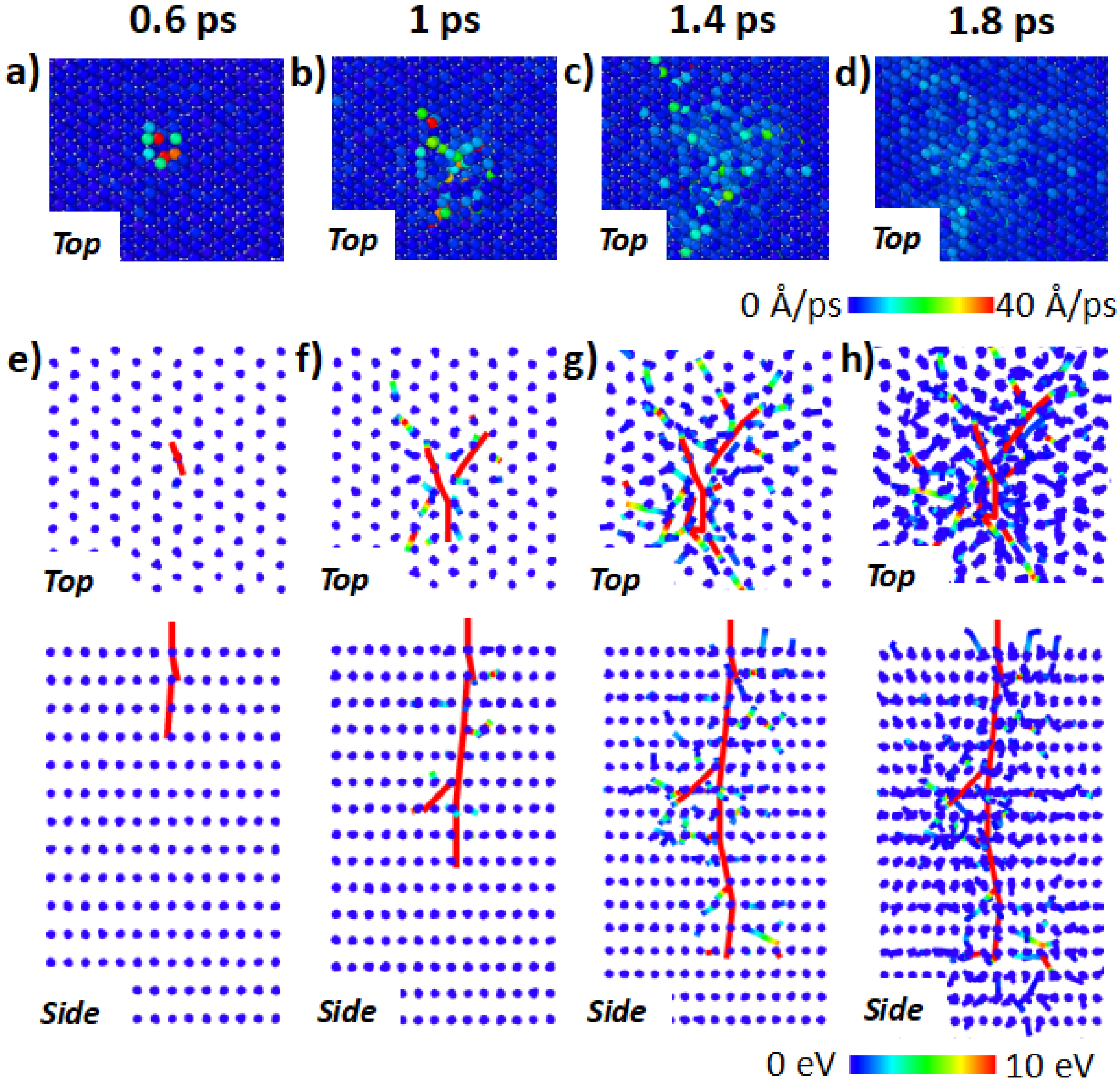

**Figure 2**. (a)-(d) Top view of atom velocity on Cu (111) surface after a 20 eV energetic Cu atom collision; (e)-(h) top and side view of atom kinetic energy distribution after the atomic collision.

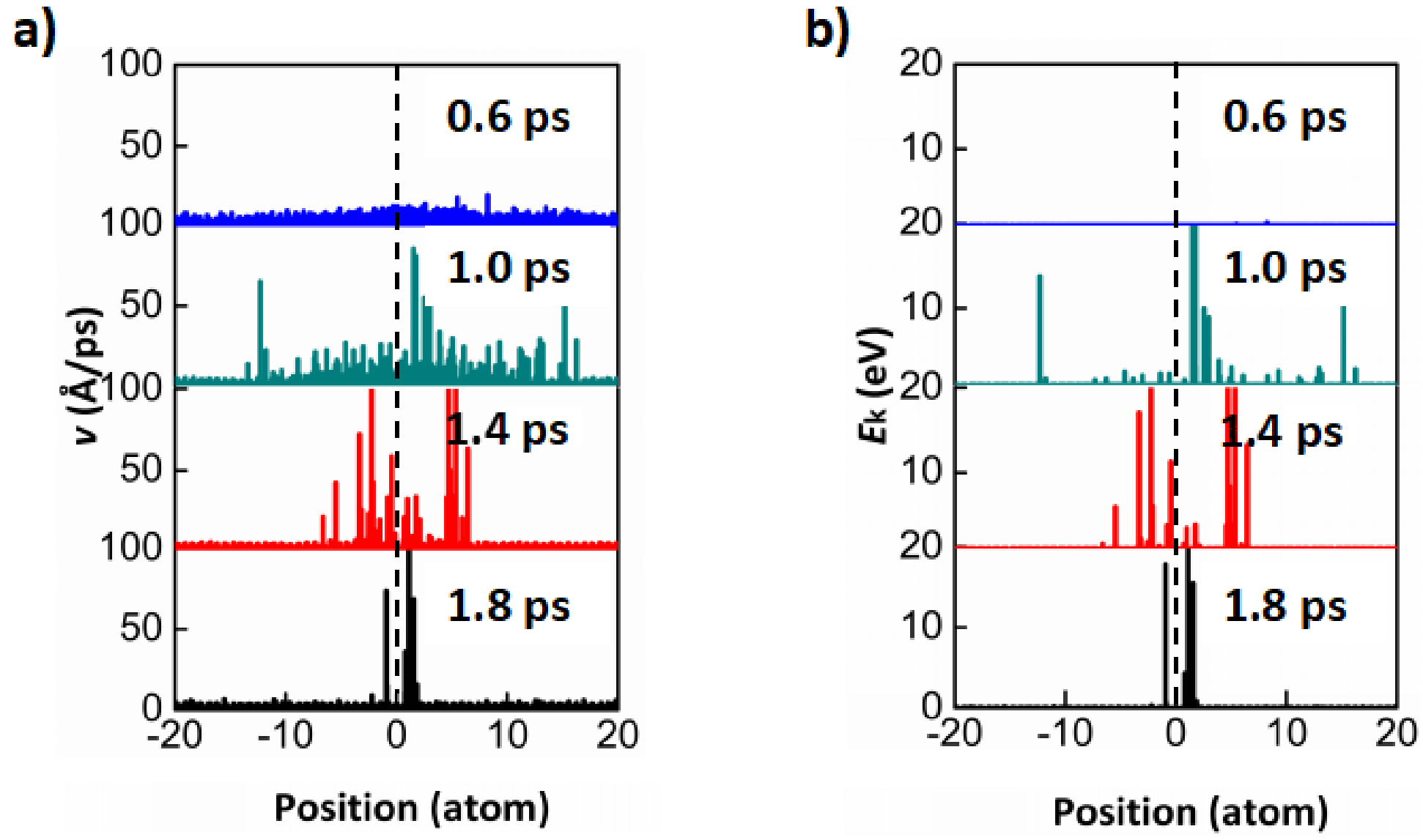

**Figure 3**. (a) Atom velocity and (b) atom kinetic energy distribution along the x-axis orientation after the atomic collision.

The deposition processes are a simplified binary collision between energetic and substrate atoms. The displacements of deposited atoms are confined within a limited range due to the dominant nuclear stopping, forming either crystalline or amorphous structure based on atomic distance. Let the energetic atoms be "hot" oscillators. In contrast, substrate atoms are "cold" oscillators, as presented in Fig. 1b and Fig. 4a. One can link these isolated oscillators using a coupled system that consists of two masses connecting by three springs, as described by PDEs (5) and (6), where $m_1$ and $k_1$ represent the mass and spring constant of "hot" oscillator, $m_2$ and $k_2$ represent the mass and spring constant of "cold" oscillator, and $k_{12}$ is the spring constant between these two coupled oscillators. Suppose each pair of neighboring oscillators is governed by the field of Morse potential $U = D_e[1 - e^{-\alpha(r-r_e)}]^2$, where $D_e$, $\alpha$, and $r_e$ are the well depth, the width of the potential, and the equilibrium bond distance, respectively, then $k_{12} = \frac{\partial^2 U}{\partial r^2} = 2\alpha^2 D_e e^{-\alpha(r-r_e)}[2e^{-\alpha(r-r_e)} - 1]$. By solving the combined equations (5) and (6), one can obtain time-dependence solutions of $u$ and $u_0$ that reflect the vibration of coupled "hot" and "cold" oscillators, respectively. Fig. 4b demonstrates a scenario of energy or amplitude exchange from the "hot" to "cold" oscillators with $r$=2 and $k$=0.1 ($k_1$=$k_2$) by a forced oscillation. The amplitude of forced oscillation towards "cold" oscillators is strikingly enhanced with a reduced atomic distance $r$ from 10 to 2, as shown in Fig. 4c and Fig. 4a, indicating the energy carried out by "hot" oscillators can be dramatically transferred to its neighboring "cold" oscillators as they are infinitely close to each other till nucleation during a binary collision. Fig. 4d presents the profile of forced oscillation with $k_1$ from 0.4 to 0.1, which emphasizes that the "hot" oscillators can make a considerable energy contribution to drive "cold" oscillators when these coupled oscillators possess the same $k$. These results highlight that energy transfer efficiency from coupled "hot" to "cold" oscillators relies on atomic distance $r$ and individual spring constant $k$. Although we've briefly discussed the energy exchange between two coupled oscillators during a binary collision, PDEs (5) and (6) cannot further describe the energy penetration among multiple phonons. We here introduce the damped wave equations as given by PDE (7), where $\beta$ and $v$ are the damping coefficient, and wave speed, respectively. If focus on the 1D solution of the heat equation (1) and the wave equation (7), as shown in Fig. 5a and Fig. 5b, respectively, a traveling spike corresponding to the maximum energy distribution exhibits along propagation direction towards the wave equation (7). Moreover, by rewriting equation (7) in the spherical coordinates, one can obtain the profiles with dramatic spikes, as depicted in Fig. 5c, and these energy spikes fundamentally contribute to localized thermal fluctuation. Fig. 5d illustrates that the energy spikes mainly depend on the wave speed within the media, and a relatively low wave speed towards the oscillation can lead to significant thermal contribution during energy deposition.

$$m_1\ddot{u}(t) + 2m_1\beta\dot{u}(t) + (k_1 + k_{12})u(t) - k_{12}u_0(t) = 0 \tag{5}$$

$$m_2\ddot{u}_0(t) + (k_2 + k_{12})u_0(t) - k_{12}u(t) = 0 \tag{6}$$

$$\ddot{u}(t,x) + \beta\dot{u}(t,x) = v^2\nabla^2 u(t,x) \tag{7}$$

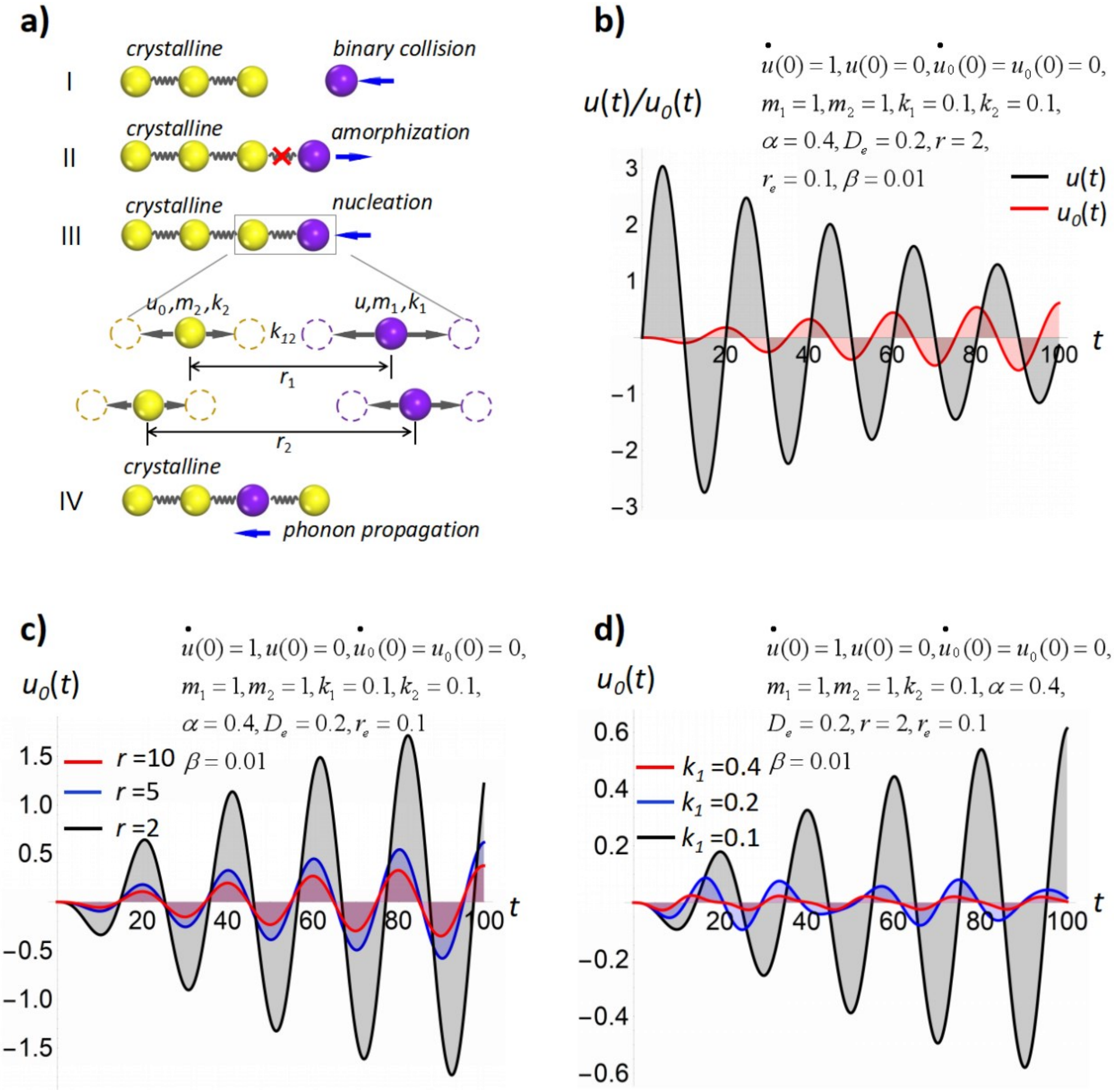

**Figure 4**. (a) Possible states among coupled-oscillators as a simplified binary collision during energetic deposition; (b) time dependence profiles of the solution towards the PDEs (5) and (6); (c)-(d) time dependence profiles of the solution towards the PDEs (5) and (6) under different *r* and *k*. The initial condition of both equations (5) and (6) is $\dot{u}(0)=1, u(0)=0, \dot{u}_0(0)=u_0(0)=0$.

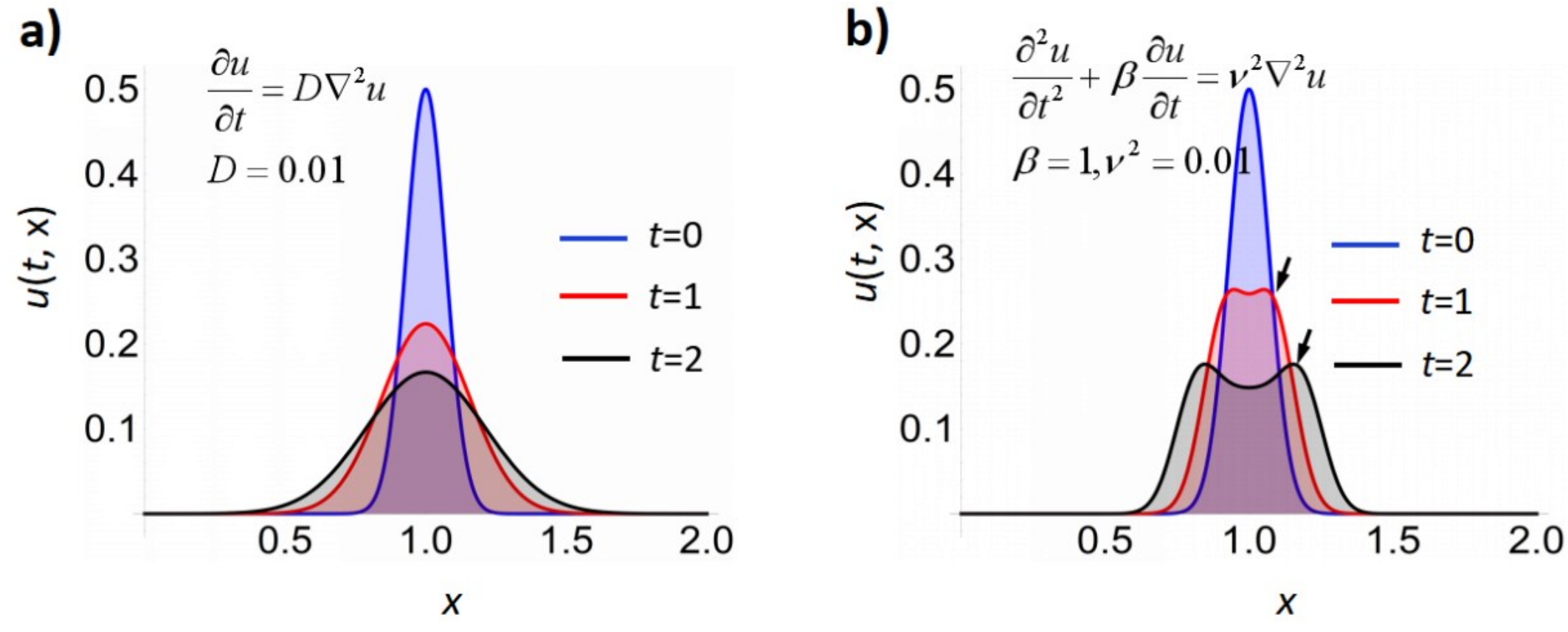

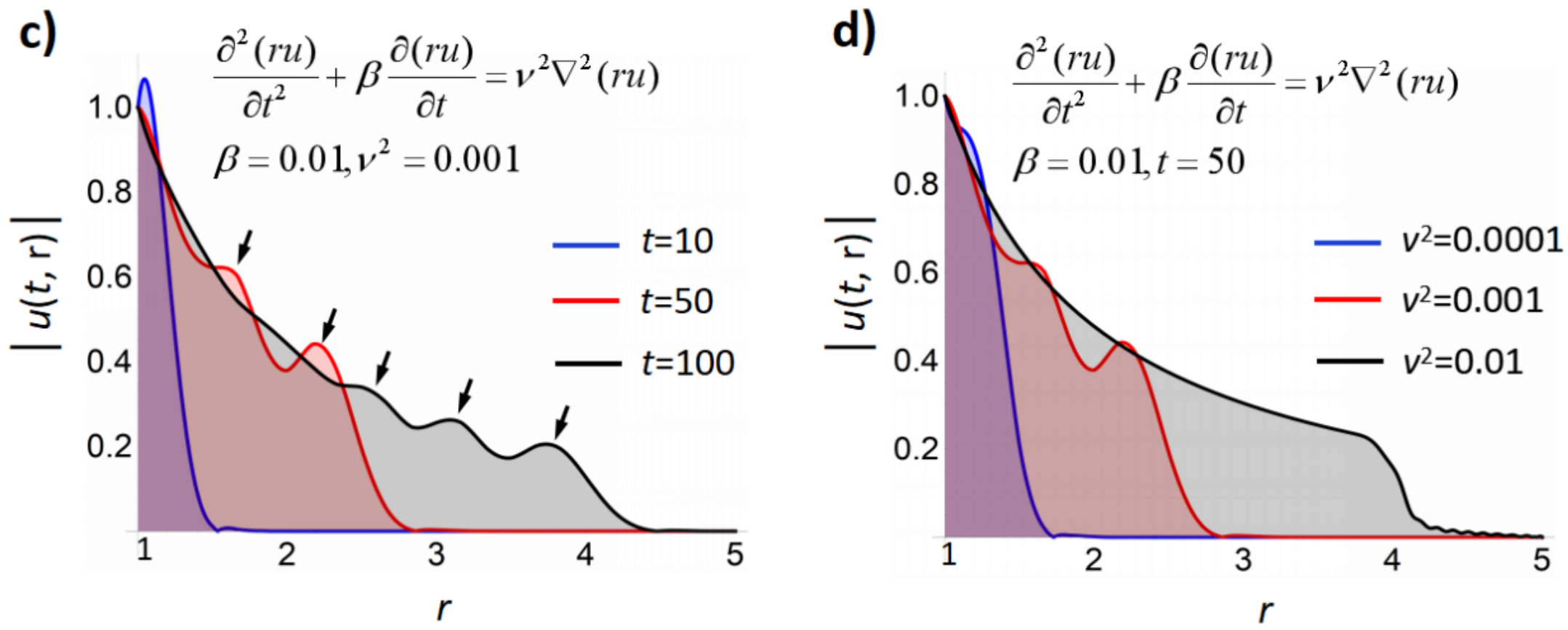

**Figure 5**. (a)-(b) Distance dependence profiles of the solution towards ordinary heat diffusion equation with the initial condition $u(t, 0)=0$, $u(t, 2)=0$, $u(0, x)=0.5e^{-100(x-1)^2}$and damped wave equation with the initial condition $u(t, 0)=0$, $u(t, 2)=0$, $u_t(0, x)=0$, $u(0, x)=0.5e^{-100(x-1)^2}$; (c)-(d) distance dependence profiles of the solution towards the damped spherical wave equation with the initial condition $u(\mathrm{t},1)=1$, $u(t, 5)=0$, $u_t(0, x)=0$, $u(0, x)=0.5e^{-100(x-1)^2}$.

## III. CONCLUSIONS

In this work, we first reproduce the thermal spike effect under the low-energy ion bombardment by performing a typical MD simulation. Then, a simplified binary collision was used to unveil the oscillation profile towards deposited atoms via coupled PDEs. The solution was essentially manifested that energy transfer efficiency from the coupled "hot" to "cold" oscillators mainly relies on the atomic distance $r$ and the oscillator spring constant $k$. The "hot" oscillators can make huge energy contributions once they are bonded to the "cold" oscillators or possess the same $k$ with the "cold" oscillators. Moreover, solving the damped wave equation can further reveal the energy propagation among coupled oscillators as the shock wave. The profiles emphasize that dramatic energy spikes will exhibit and contribute to localized thermal fluctuation with a relatively low wave speed $v$ within the media, which provides a more profound understanding of the thermal spike effect than the normal heat diffusion equation.

## ACKNOWLEDGMENTS

This work is partly supported by National Natural Science Foundation of China (No.12175019), National Natural Science Foundation Joint Fund Key Project (U1865206), and China scholarship council (CSC).